\documentstyle[prl,aps,preprint]{revtex}
\begin{document}

\title{Self similar Barkhausen noise in magnetic domain wall motion.}

\author{Onuttom Narayan}
\address{University of California at Santa Cruz, Santa Cruz, CA 95064}

\date{\today}

\maketitle

\begin{abstract}
A model for domain wall motion in ferromagnets is analyzed.
Long-range magnetic dipolar interactions are shown to give
rise to self-similar dynamics when the external magnetic
field is increased adiabatically. The power spectrum of 
the resultant Barkhausen noise is of the form $1/\omega^\alpha$,
where $\alpha\approx 1.5$ can be estimated from the critical
exponents for interface depinning in random media.
\end{abstract}

\pacs{}

When a domain wall in a ferromagnet moves in response to
a change in the externally applied magnetic field, it is
known to do so in a jerky, irregular manner. As a result
of this irregular motion,
the magnetization changes in bursts, leading to the phenomenon
of Barkhausen noise.
The reason for the unevenness in the motion is that the domain
wall is pinned in various places by impurities in the material.
The domain wall moves forward by breaking free of the impurities 
holding it back, only to be obstructed by impurities further ahead.
A simple model for the dynamics has been proposed, in 
which the coordinate of the 
domain wall is treated as a single dynamical variable.~\cite{Alessandro}
As one might expect, within such a model the temporal 
fluctuations in the motion of the domain wall (revealed in the 
Barkhausen noise) yield information about the spatial distribution
of the impurities in the material. However, more recent experiments 
have revealed that this single degree of freedom model for domain wall
motion is essentially incomplete.~\cite{Weiss}

The reason for the inadequacy of the model is that a magnetic
domain wall is a {\it spatially extended\/} object, with a large 
number of degrees of freedom. Under slow driving, the dynamics
of the domain wall are expected to be governed by the collective
behavior of these multiple degrees of freedom. This is 
reminiscent of ``depinning transitions'' seen in a variety of
driven systems, where close to the transition the dynamics
are affected qualitatively by collective behavior.~\cite{depinn}

Despite the similarities, there is an important difference 
between magnetic domain wall motion and conventional depinning
transitions. For any value of the external driving force (the
magnetic field), a magnetic domain wall reaches a stationary
configuration. This stationary configuration appears to be 
self-similar, a fact inferred experimentally from the power-law
correlations in the 
Barkhausen noise generated when the magnetic field is slowly
increased. This is in sharp contrast to conventional depinning
transitions, where increasing the external force results
in a phase transition from a static to a moving phase at a 
critical force, and self-similarity is seen only at the
transition. The apparent self-similarity in magnetic domain 
walls achieved without specially adjusting the external 
magnetic field is reminiscent of the concepts of self-organized
criticality.~\cite{btw}

In a recent paper, Urbach {\it et al\/}~\cite{Urbach} 
provide evidence that this departure from conventional
depinning transitions is caused by the presence of 
long-range dipolar interactions in a ferromagnet. These
dipolar interactions push the domain wall towards the 
center of the system. In addition, they also produce 
long-range effective forces between different 
parts of the domain wall. Urbach {\it et al\/} numerically
solve a model for the dynamics
with an approximate treatment of these long-range 
forces.~\cite{Urbach} In one limit, the magnetic force
is taken to be infinite-ranged, and the numerics indeed 
yield a power-law distribution for the power spectrum of the
resultant Barkhausen
noise. However, in the opposite limit, where the magnetic
force is taken to be local, self-similar behavior is not seen.

Thus while the tendency of the interactions to push the 
domain wall towards the center of the system is sufficient to 
destroy the moving phase commonly seen in such externally driven systems
(and thereby the depinning 
transition leading to it), the exact nature of the interactions
is important in 
determining whether the resultant state is self-similar or
not. It is not clear whether an accurate description of the 
forces induced by the dipolar interactions, which must lie
between the two limits considered by Urbach 
{\it et al\/},~\cite{Urbach} will result
in self-similar behavior. In this paper I analyze 
the dynamics of a magnetic domain wall without 
any approximations for the dipolar interactions, 
verifying that the resultant behavior is indeed self-similar.

Following Urbach {\it et al\/},\cite{Urbach} I consider 
a two dimensional Ising system magnetized perpendicular 
to the plane. A single domain wall is assumed to run approximately
parallel to one of the sides of the system (the transverse direction),
and close to its midpoint. The domain wall is characterized by
its (small) longitudinal displacement $h(x,t)$ as a function of the 
transverse coordinate $x$ and time $t$. (In $d$ dimensions, 
$x$ is generalized to a $(d-1)$ dimensional vector.) The 
equation of motion used for the motion of the domain wall is
very similar to the one used by Urbach {\it et al\/}:~\cite{Urbach}
\begin{equation}
\Gamma\partial_t h(x,t)=k\partial_x^2 h(x,t)+u(h(x,t);x)+H
+\int dx^\prime I(x,x^\prime) h(x^\prime,t).\label{eom}
\end{equation}
This equation is obtained by neglecting inertial effects and
thermal noise, so that the dynamics are purely relaxational.
$\Gamma$ is a constant characterizing the amount of dissipation.
The surface tension of the domain wall gives rise to the first 
term on the right hand side, the next term is due to the pinning
forces from the impurities, and $H$ is the applied magnetic field.
The last term in the equation is obtained by expanding the dipolar interaction
energy to second order in $h(x,t)$ as $-{1\over 2}
\int dx dx^\prime I(x,x^\prime)h(x,t)h(x^\prime,t)$,
and differentiating with respect to $h$. 

We now evaluate the last term in Eq.(\ref{eom}) above. For the 
perpendicular Ising system considered here, the interaction 
between two dipoles at ${\bf r}_1$ and ${\bf r}_2$ is isotropic,
and proportional to $|{\bf r}_1-{\bf r}_2|^{-3}.$ Translational 
invariance requires that (neglecting edge effects) $I(x,x^\prime)=
I(x-x^\prime)$. From the scale invariance of $|{\bf r}_1-{\bf r}_2|^{-3},$
and power counting, the dipolar interaction energy expanded to 
second order in $h$ must be of the form 
$-{1\over 2}\int dq h(q)h(-q)[f(qL)/L^2]$, where
$L$ is the linear extent of the system. By considering a uniform
displacement, $h(x)$ independent of $x$, it is possible to verify
that $f(qL)$ has a finite $q=0$ limit. Thus the $q=0$ limit yields
an effective restoring force that drives the domain wall towards the 
center of the system ($h=0$), while for $qL>>1$ one obtains an 
effective reduction in the 
surface tension $k$, which has no qualitative effect.

If the applied external magnetic field increases at a slow constant rate,
the domain wall moves forward at, on the average, a constant
rate. The deviation from this uniform motion obeys an equation that
can be obtained from Eq.(\ref{eom}).
Apart from the last term, the resultant equation
is the same as for conventional interface depinning.\cite{interf}

The crucial feature of the restoring force obtained from the dipolar
interactions is that it 
is {\it small\/} for large systems. Thus although the 
$[kq^2+f(qL)/L^2]h(q)$ term in Eq.(\ref{eom}) has a `mass term', and
therefore a cutoff to the self-similar behavior, this cutoff diverges
with system size. In fact, since $1/L^2$ scales in the same manner
as $q^2$ under the renormalization group (and neither term receives
loop corrections),\cite{interf} the only cutoff to the scaling of 
quantities like
$\langle h(q)h(-q)\rangle$ is the standard finite size cutoff. 
(There is another cutoff if the external magnetic field is 
increased at a finite rate instead of adiabatically.)

The scaling of the wavelength and frequency dependent fluctuations in
$h$ for the domain wall is obtained from the corresponding
interface result:
\begin{equation}
\langle h(q,\omega)h(-q,-\omega)\rangle=
L q^{-(2\zeta+1+z)}\Phi(qL,q^z/\omega),\label{corrln},
\end{equation}
where $\zeta$ is the 
roughness exponent of the interface and $z$ the dynamic exponent. 
(This form can be seen to be correct by integrating over $\omega$,
and expressing $\langle h(q,t)h(-q,t)\rangle$ in terms of 
$\langle h(x,t)h(x^\prime,t)\rangle$ which scales as $|x-x^\prime|
^{2\zeta}$.)
The Barkhausen noise measures the temporal fluctuations in the 
rate of change of the total magnetization of the system. These 
are proportional to the fluctuations in the spatially averaged
velocity of the domain wall. Thus the power spectrum of the Barkhausen
noise is obtained from the velocity-velocity correlation function
of the domain wall. Multiplying the right-hand side of Eq.(\ref{corrln})
by $\omega^2$ and taking the $q\rightarrow 0$ limit yields 
for the correlations in the spatially averaged velocity
\begin{equation}
L^{-2}\langle {\partial_t h}(\omega){\partial_t h}(-\omega) \rangle
=F(\omega L^z)/[L\omega^{(2\zeta+1)/z-1}].
\end{equation} 
Ignoring the finite 
size cutoff which occurs at very low frequencies, this has a power-law
form. Numerical estimates for the 
critical exponents in two dimensions\cite{estimate} indicate that 
$(2\zeta+1)/z-1$ is close to 1.5. 
This is lower than the exponent of 2 for the power spectrum
of the Barkhausen noise obtained in early experiments,\cite{exp} 
but is not inconsistent with recent experimental results.\cite{Baggins} 

It is clear why the short-range model of Urbach {\it et al\/} did
not yield self-similar behavior, since the mass term in the equation
of motion (and thus the cutoff) is finite. The mean field limit that
they consider is even
simpler: since $f(qL)=0$ for $q\neq 0$, expressing Eq.(\ref{eom}) in terms
of $h(x,t)-\overline h(t)$ results in exactly the same equation
as for conventional interface depinning. The interface velocity is replaced
by the domain wall velocity. Since this tends to zero when the external
magnetic field is increased adiabatically, the system is at its critical point.

The analysis above is easily generalized to other dimensions, since in
all dimensions the interaction energy has a scaling form and a $q=0$
limit proportional to $1/L^2$.  The power law for the 
Barkhausen noise power spectrum 
will however be different in different dimensions, most
notably for $d=1$, where the exponent of the power law should be zero. This
is in contrast to the single degree of freedom model originally proposed 
for the dynamics,\cite{Alessandro}
where a $1/\omega^2$ dependence is predicted independent of dimension. 
Extending the results obtained above
to systems with multiple domain walls remains an open issue.

I thank Karin Dahmen, Mehran Kardar, Jim Sethna, Jeff Urbach and Mike 
Weissman
for useful discussions. This work was partly completed at the 
Aspen Institute for Physics and supported by the A.P. Sloan
Foundation.


\begin{references}
\bibitem{Alessandro} B. Alessandro, C. Beatrice, G. Bertotti and A. Montorsi,
J. Appl. Phys. {\bf 68}, 2901 (1990).
\bibitem{Weiss} K.P. O'Brien and M.B. Weissman, Phys. Rev. E{\bf 50},
3446 (1994).
\bibitem{depinn} O. Narayan and D.S. Fisher, Phys. Rev. B{\bf 46}, 11520
(1992); N. Martys, M. Cieplak and M.O. Robbins, Phys. Rev. Lett. {\bf 66};
D. Ertas and M. Kardar, Phys. Rev. E{\bf 49}, R2535 (1994).
\bibitem{btw} P. Bak, C. Tang and K. Wiesenfeld, Phys. Rev. Lett. {\bf 59},
381 (1987).
\bibitem{Urbach} 
J.S. Urbach, R.C. Madison and J.T. Markert, Phys. Rev. Lett. {\bf 75}, 276
(1995).
\bibitem{interf}
O. Narayan and D.S. Fisher, Phys. Rev. B {\bf 48}, 7030 (1993).
\bibitem{estimate} H. Leschhorn, T. Nattermann, S. Stepanow and 
L.-H. Tang, Annalen der Physik (in press).
\bibitem{exp}W. Grosse-Nobis, J. Magn. and Magn. Mat. {\bf 4}, 247 (1977).
\bibitem{Baggins} G. Durin, G. Bertotti and A. Magni, Fractals {\bf 3}, 
351 (1995).
\end{references}
\end{document}